\documentclass[useAMS,usenatbib,usegraphicx]{mn2e}

\usepackage{psfig}
\usepackage{times}
\usepackage{subfigure}
\usepackage{lscape}

\newcommand{\rsun}{\,\mbox{$R_{\odot}$}}
\newcommand{\rjup}{\,\mbox{$R_J$}}

\newcommand{\mjup}{\,\mbox{$M_J$}}

\newcommand{\kms}{\hbox{km s$^{-1}$}}

\newcommand{\degs}{$\degr$}
\newcommand{\chisq}{$\chi^{2}$}

\title[HD 189733b 2.2\micron\ contrast ratio]{Limits on the 2.2\micron\ contrast ratio of the close orbiting planet \hbox{HD 189733b}}

\author[J.R.~Barnes et al.]
{J.R.~Barnes$^1$\thanks{E-mail: j.r.barnes@herts.ac.uk} 
 Travis~S.~Barman$^2$,
 L.~Prato$^2$,
 D.~Segransan$^3$,
 H.R.A.~Jones$^1$, 
 C.J.~Leigh$^4$, 
 \newauthor
 A.~Collier~Cameron$^5$ and
 D.J.~Pinfield$^1$ \\
$^1$ Centre for Astrophysics Research, University of Hertfordshire, Hertfordshire AL10 9AB. UK. \\ 
$^2$ Lowell Observatory, Planetary Research Centre, 1400 West Mars Hill Road, Flagstaff, AZ 86001. USA \\
$^3$ Observatoire de Gen\`{e}ve, 51 Chemin des Maillettes, 1290 Sauverny. CH. \\
$^4$ Astrophysics Research Institute, Liverpool John Moores University, Birkenhead CH41 1LD. UK. \\
$^5$ SUPA, School of Physics and Astronomy, University of St Andrews, Fife KY16 9SS. UK. \\
}

\begin{document}

\date{Accepted 2007 August 27. Received 2007 August 24; in original form 2007 July 20}


\maketitle

\label{firstpage}

\begin{abstract}

We obtained 238 spectra of the close orbiting extrasolar giant planet \hbox{HD 189733b} with resolution $R \sim 15,000$ during one night of observations with the near infrared spectrograph, NIRSPEC, at the Keck II Telescope. We have searched for planetary absorption signatures in the \hbox{$2.0$\ \micron\,-\,$ 2.4$\ $\micron$} region where H$_2$O and CO are expected to be the dominant atmospheric opacities. We employ a phase dependent orbital model and tomographic techniques to search for the planetary absorption signatures in the combined stellar and planetary spectra. Because potential absorption signatures are hidden in the noise of each single exposure, we use a model list of lines to apply a spectral deconvolution. The resulting mean profile possesses a S/N ratio that is 20 times greater than that found in individual lines. Our spectral timeseries thus yields spectral signatures with a mean S/N = $2720$. We are unable to detect a planetary signature at a contrast ratio of \hbox{log$_{10}$($F_{p}/F_{*}) = -3.40$}, with 63.8 per cent confidence. Our findings are not consistent with model predictions which nevertheless give a good fit to mid-infrared observations of \hbox{HD 189733b}. The 1$-\sigma$ result is a factor of 1.7 times less than the predicted \hbox{2.185 \micron}\ planet/star flux ratio of \hbox{log$_{10}$($F_{p}/F_{*}) \sim -3.16$}.

\end{abstract}


\section{Introduction}
\protect\label{section:intro}

A factor governing the atmospheric physics of the gas giants in our own solar system is the relatively low temperatures of the surrounding environment. While internal heat sources and rotation play a role in atmospheric dynamics, much of the structure seen in the atmospheres is governed by chemical species which appear in relatively low abundance ($\sim 1$ per cent by mass) compared with a dominant makeup of Hydrogen and Helium. Extra-solar planetary atmospheres amenable to study must however differ considerably from the solar system paradigm on account of their close proximity to their parent star.

Modern instrumentation is now capable of making observations of Close-orbiting Extrasolar Giant Planets (CEGPs) in order to learn about the physical make up of their atmospheres. Reflected light studies \citep{cameron99tauboo,charbonneau99tauboo,cameron02upsand,leigh03a,leigh03b} and more recent MOST photometry \citep{rowe06} results have placed albedo {\em upper limits} on the atmospheres of CEGPs in the optical. While reflected light studies only search for an attenuated copy of the stellar light, they have the potential to yield important information about the chemical makeup of the planetary atmospheres. These upper limits to the planet/star contrast ratio have only been able to constrain CEGP atmospheres as having low reflectivity when compared with the solar system gas giants.

The contrast between the planet and star in the infrared offers a more favourable regime in which to study CEGPs. An estimate of the temperature of a CEGP was first reported from 24 \micron\ secondary eclipse measurements of \hbox{HD 209458b} by \citet{deming05hd209458}. While a number of subsequent mid-infrared observations of other systems have now been made with the Spitzer Space Telescope, \citet{harrington06} were the first to report measurement of the phase dependent flux of a CEGP, implying a world with a significant day-night temperature gradient. Such observations are now fuelling new research into the meteorology of extrasolar planets \citep{cooper05,cooper06,fortney06}. To obtain a fuller understanding of the atmospheres of CEGPs it is important to study as wide a region of wavelength space as possible on CEGPs with a range of parameters.

The detection of narrow absorption features due to molecular species is difficult in the optical because the atmospheres appear so dark whereas no high resolution instrumentation in the thermal infrared is widely available. A number of contemporary instruments such as NIRSPEC at Keck \citep{mclean98nirspec}, IRCS at Subaru \citep{kobayashi00ircs} and CRIRES at VLT \citep{kaufl04crires} are capable of making searches for absorption features in the near infrared possible.

\begin{table*}
\caption{NIRSPEC/Keck II observations of \hbox{HD 18733} and \hbox{HD 192538} for UT 2006 July 22. Cloud cover was highly variable throughout the night. Groups of 15 spectra from the first \hbox{HD 189733} block were coadded post-observations to give the same effective exposure time (60 secs) as in the second block, resulting in an effective total of 238 frames. As a result of variable cloud cover, we were unable to extract 13 of the \hbox{HD 189733} spectra, leaving 225 usable frames.}
\protect\label{tab:journal}
\vspace{5mm}
\begin{center}
\begin{tabular}{lcccccl}
\hline
Object			& UT start of	& UT start of	& Time per	  & Number of		& Number of 	& Comments  \\
			& first frame	& last frame	& exposure [secs] & coadds per frame	& observations	& 	    \\
\hline
\multicolumn{7}{c}{UT 2006 July 22} \\
\hline
HD 192538		& 05:48:46	& 05:59:18	&    	60	  &	1		& 8		& A0V template	      \\
HD 189733		& 06:10:58	& 06:43:52	&	4	  & 	1		& 125		& Software crash at end of sequence			 \\
HD 189733		& 07:04:27	& 14:08:0922	&	4	  & 	15		& 230		& 			\\

\hline
\end{tabular}
\end{center}
\end{table*}


\subsection{HD 189733}
\protect\label{section:hd189733}

HD 189733b was the first planet to be discovered as a transiting planet spectroscopically \citep{bouchy05hd189733}. Observations taken around the time of transit indicated radial velocity anomalies due to the Rossiter-McLaughlin effect, where the transiting planet occults part of the stellar disc leading to a deviation from the measured Keplerian solution. Further spectroscopic observations were used to determine the spin-orbit alignment \citep{winn06}, indicating that the sky projections of the stellar spin axis and orbit normal are aligned to with \hbox{1.4\degs $\pm$ 1.1\degs}. \citet{bakos06} used astrometric and radial velocity measurements to show that \hbox{HD 189733} is the \hbox{(K1V\,-\,K2V)} primary \citep{bouchy05hd189733} in a double star system with the secondary mid-M dwarf companion orbiting at 216 AU from \hbox{HD 189733} in a plane orthogonal to the orbit of \hbox{HD 189733b}.

The system parameters derived in the discovery paper by \citet{bouchy05hd189733} have since been refined through inclusion of multiband photometric transit observations by \citet{bakos06hd189733refine}. More recently, \citet{winn07hd189733b} have also used photometric observations out of eclipse to determine a stellar rotation period of 13.4 d. A total of eight transits were used to determine the stellar and planetary radii and the photometric mid-transit ephemeris of \hbox{H189733b} which are $R_* = 0.753 \pm 0.025$\rsun, $R_p = 1.156 \pm 0.046$\rjup, and $T_c = 2453988.80336(23) + 2.2185733(19)$\ d\ respectively (where the figures in brackets for the ephemeris denote the uncertainty in the last two decimal places). The values of $95.79 \pm 0.24$\degs\ \citep{bakos06hd189733refine} and $95.76 \pm 0.29$\degs\ determined for the inclination of the orbit of \hbox{HD 189733b} are in good agreement and remove the ambiguity in the planet's mass, yielding $Mp = 1.13 \pm 0.03$\mjup.

Because \hbox{HD 189733b} is one of the closest orbiting planets (\mbox{$a = 0.0312 \pm 0.0004$}AU) with a relatively late spectral type parent star , one would expect a relatively high planet/star contrast ratio, $F_p/F_*$. \citet{deming06hd189733b} detected the strong infrared thermal emission in the 16 \micron\ band using the Spitzer space telescope. Photometry enabled a clear detection of the eclipse with a depth of $0.551 \pm 0.030$ per cent, or $F_p/F_* = 0.0055 \pm 0.0003$, indicating the star to be only 181 times brighter than the planet at 16 \micron\ near to eclipse. 

\citet{tinetti07} modelled transmission spectra, which are dominated by H$_2$O and CO opacities, and presented simulations to show that these features should be deep enough to produce signatures which are detectable in the thermal infrared by present and future space based missions. { \citet{tinetti07nature} have subsequently used planetary transit depth observations made with the Spizter Space Telescope Infrared Array Camera (IRAC) at 3.6 \micron, 5.8 \micron\ \citep{beaulieu07} and 8 \micron\ \citep{knutson07hd189733b} in conjunction with transmission spectra modelled using H$_2$O opacity lists \citep{barber06water} to infer the presence of water in the atmosphere of \hbox{HD 189733b}. \citet{grillmair07spectrum} however have reported  measurement of the 7.5\,-\,14.7 \micron\ spectrum of \hbox{HD 189733b} using the Spizter Space Telescope Infrared Spectrograph (IRS) which does not show the expected decrease \citep{burrows06} in flux at the short wavelength end of this range due to increasing H$_2$O opacity. It has been suggested \citep{fortney07water} that the discrepancy between the IRAC 8 \micron\ flux and IRS flux may result from unreliable reduction of the IRS spectra and that the IRAC photometry is thus in good agreement with one dimensional atmospheric models of \hbox{HD 189733b} in the 8 \micron - 16 \micron\ range spanned by IRAC flux measurements.}

\citet{knutson07hd189733b} observed 0.12 per cent flux variation between transit and secondary eclipse, with a maximum 16\degs\ before secondary eclipse. Their \hbox{8 \micron}\ Spitzer observations were used to derive a simple surface map showing a hot spot offset by 30\degs\ East of the substellar point.

In this paper we aim to detect the absorption signature of \hbox{HD 189733b} at $\sim 2.2$ \micron. The method outlined in \S \ref{section:search} has the potential to determine the planet/star contrast ratio, $F_p/F*$, if an absorption signature is detected \citep{barnes07hd75289}. The detection of such opacities forms an important test for the presence of atmospheric molecular species.

\section{Observations and Data Reduction}

Observations were made with NIRSPEC \citep{mclean98nirspec} at the Keck II Telescope on UT 2006 July 22. A high cadence timeseries of \hbox{HD 189733} spectra were recorded using a 1024x1024 InSb Aladdin-3 array. The rapidly rotating bright A0V star, \hbox{HD 192538}, was observed at the start of the night to enable identification of telluric features in the spectra. With the NIRSPEC-7 blocking filter, a wavelength span of 2.0311 \micron\,-\,2.3809 \micron\ was achieved at a resolution of \hbox{R $\sim$ 15000}. The observations are summarised in Table \ref{tab:journal}.

\subsection{Data extraction}
\protect\label{section:extract}

Each raw frame was corrected by subtracting a scaled dark current frame. Pixel to pixel variations were removed using flat-field exposures taken with an internal tungsten reference lamp. In order to create a reliable balance frame to remove the pixel sensitivity variations, we divided a Gaussian blurred (using a FWHM of 7 pixels) version of the master flat field image by the original master flat field image. Two flat fields each comprising of 100 median-combined frames taken at both the start and end of the night were used since a software crash at the end of the first sequence of frames listed in Table \ref{tab:journal} resulted in an instrument reset. After reboot a small shift of 2 pixels in wavelength position of the grating occurred. Examination of the balance frames shows that while blemishes and hot pixels remain unchanged, a striped or interference pattern present in the balance frame taken at the start of the night has shifted in the balance frame taken at the end of the night. The interference pattern varies from order to order and repeats with a length scale of between 10 and 20 pixels. If this pattern is purely dependent on the light path, it should largely be removed as a result of flatfielding. By dividing one balance frame by the other, and measuring the residual in each order, we estimate the flatfield stability limited by shifts in the interference pattern after instrument setup changes (2 pixel shift) through the night to be 1.0 per cent. The flatfield at the start of observations was used for the first block of observations which only comprised 8 sets of co-added frames (Table \ref{tab:journal}, see caption), while the flatfield taken at the end of observations was used in conjunction with the remainder of observations after the instrument was reset.

The worst cosmic ray events were removed at the pre-extraction stage using the Starlink {\sc figaro} routine {\sc bclean} \citep{shortridge93figaro}. The recorded frames comprised 4 sec exposures at the start of the night before the software crash (see Table \ref{tab:journal}) and \hbox{$15 \times 4$\ secs\ $= 60$\ secs} exposures thereafter. We therefore coadded groups of 15 exposures from the first set of frames to ensure that all spectra were exposed for the same \hbox{60 secs}.

A number of sky lines were present in the four shortest wavelength orders while the two reddest orders contained only one very weak line. In order to maximise the stability of observations made with the instrument, we decided against using a nodding ABBA sequence. We were thus unable to use subsequent A-B and B-A differences to reject sky lines during the extraction. We followed a similar method to that detailed in \citet{barnes07hd75289}, finding that the Starlink \'{e}chelle data reduction package, {\sc echomop} \citep{mills92}, rejects all but one strong sky line, at \hbox{$\sim21501.5$ \AA}\ through iterative fitting of a third degree polynomial. Bad pixels on the detector close to the location of this strong sky line meant that a small region encompassing the line was excluded from further analysis. The spectra were extracted using {\sc echomop}'s implementation of the optimal extraction algorithm developed by \citet{horne86extopt}. {\sc echomop} propagates error information based on photon statistics and readout noise throughout the extraction process. \\

\subsection{Wavelength calibration}
\protect\label{section:wavcalib}

Calibration frames using a combination of Ar, Ne, Xe and Kr lamps were taken during the observations although the line identifications are poor in some orders. We therefore adopted the approach used in \citet{barnes07hd75289} where telluric features in the observed spectrum of the standard star were used. A spectrum generated from a HITRAN line list \citep{rothman05hitran} enabled identification of the corresponding features with known/measurable wavelengths. Between 10 and 19 lines were used in each order and a cubic polynomial was used to give fits with a rms scatter equal to \hbox{$1.44 \times 10^{-6}$ \micron}\ (0.0144\ \AA). The mean wavelength increment of the six calibrated orders (NIRSPEC orders 32\,-37) is \hbox{$3.23 \times 10^{-5}$ \micron}\ (0.323 \AA). At the 2.185 \micron\ centroidal wavelength position of our deconvolved lines (see \S \ref{section:decon}) the rms scatter in the fit corresponds to 0.045 pixel elements or 0.00099 resolution elements. The wavelength ranges for each of the six orders are: \hbox{2.031\,-\,2.061 \micron} (order 37), \hbox{2.087\,-\,2.118 \micron} (order 36), \hbox{2.146\,-\,2.178 \micron} (order 35), \hbox{2.209\,-\,2.242 \micron} (order 34), \hbox{2.275\,-\,2.309 \micron} (order 33), \hbox{2.346\,-\,2.381 \micron} (order 32).


\begin{figure*}
\begin{center}
\vspace{-3.0cm}
\hspace{-2.0cm}
\vbox to270mm{\includegraphics[height=260mm,angle=180]{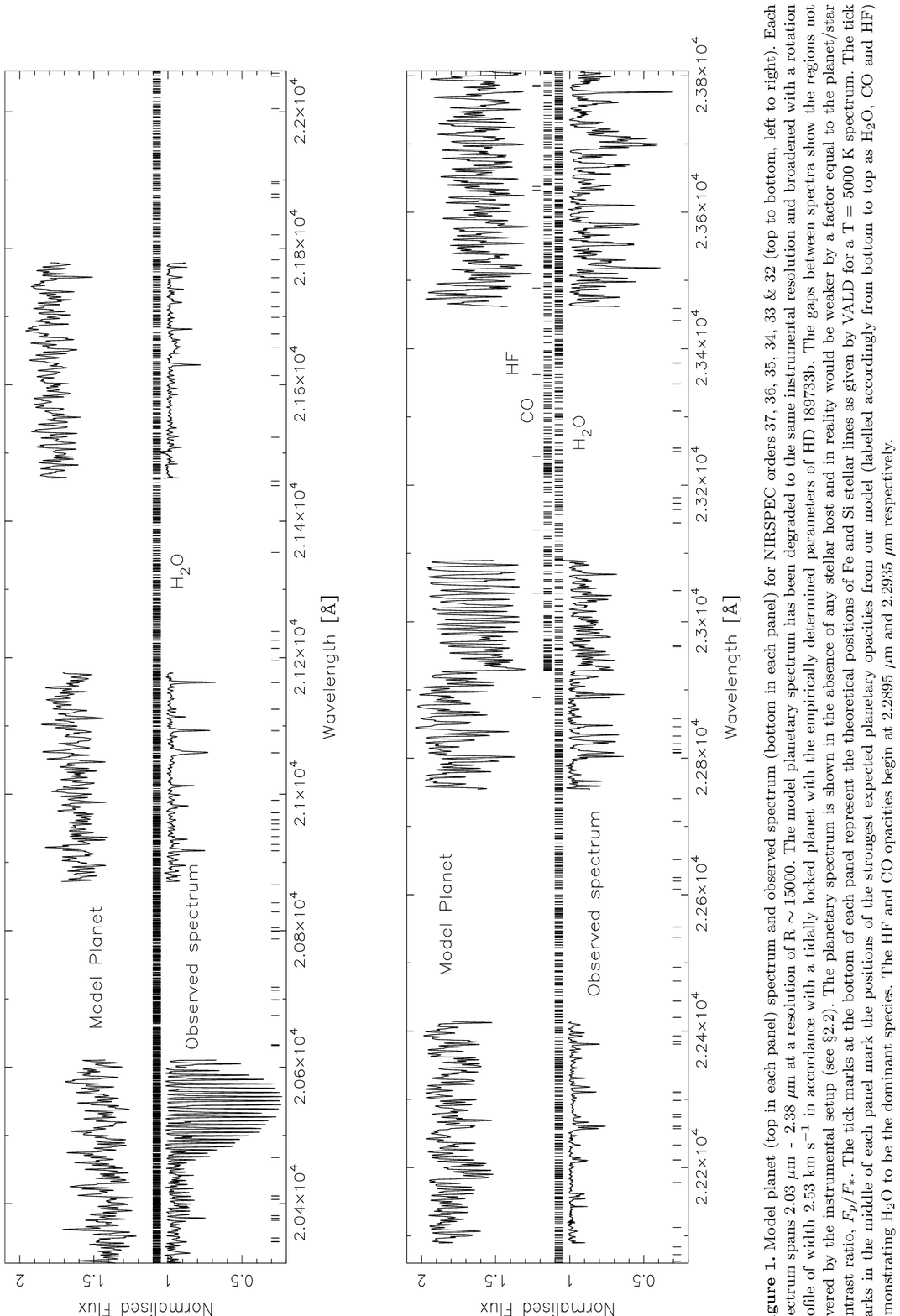}
\caption{}
\vfil}
\end{center}
\protect\label{fig:planet_spectrum}
\end{figure*}

\section{Searching for a planetary signature}
\protect\label{section:search}

\subsection{Removal of stellar and telluric spectra}
\protect\label{section:removespec}

Before searching for a planetary signal, it is important to remove the stellar and telluric lines from our spectra. This is achieved by subtracting a shifted and scaled template spectrum that has been created by coadding all the frames from our night of observations. This procedure is valid because the planetary signal is not stationary during the observations. The phases of the observations made here mean that any planetary absorption signature will be Doppler shifted from \hbox{+60 \kms}\ to \hbox{-80 \kms}\ during the planet's orbit about \hbox{HD 189733}. The template spectrum was fitted to each \hbox{HD 189733} spectrum in turn by taking derivatives of the spectra and using splines to calculate the scale factor at points across the spectra. This process can account for those lines that behave independently over the night and is described in \citet{cameron02upsand} (Appendix A). For example, all telluric lines do not necessarily vary in strength by the same factor at any given time while the strength of all telluric lines may vary relative to the stellar lines. Fig. 1 (tick marks at bottom of each panel) shows the positions of stellar lines (for further details see \S \ref{section:decon}) from the Vienna Atomic Line database \citep{kupka99}. A comparison with the observed template spectrum above the tick marks shows that the spectra are dominated by the telluric lines. 

While this template subtraction procedure is effective at removing the stellar and telluric lines and preserving a possible planetary signal, some residuals remain in the spectra. These residuals often appear in blocks of spectra and become evident when each echelle order is displayed in a greyscale image. The large variability in S/N ratio of the current spectra (\S \ref{section:decon}), due to highly variable cloud cover, prompted us to investigate removal of the blocks of spectra containing the fewest counts. In \S \ref{section:results} we report our findings for the case where 42 per cent of the spectra were rejected prior to further analysis.

Remaining systematics were removed using principal components analysis as described by \citet{cameron02upsand} (Appendix B); and discussed at length in \citet{barnes07hd75289}. A balance must be struck between removal of systematics and removal of photons which would reduce our detection threshold. Only residuals at a fixed wavelength position are removed if the first few principal components are subtracted from the data, ensuring a moving planetary signal is not significantly attenuated. We subdivided each of the six orders into three subsections and subtracted the strongest components (between 2 and 5) for each section.

\subsection{Atmospheric Model}
\protect\label{section:atmos}

{ We attempt to make optimal use of the information content of our spectra by considering all the planetary absorption lines contained within the \hbox{$\sim 1645$}\ \AA\ wavelength span of our data. We therefore require a model that describes the planetary atmosphere as accurately as possible. \citet{barnes07hd75289} discuss the nature of the irradiated model atmosphere that includes day-night time temperature gradients on the planet. For a full description of the model opacities and setup see \citet{ferg05}, \citet{barman01} and \citet{bha05} BHA05. For \hbox{HD 189733b}, we model an atmosphere with solar metallicity for a temperature, \hbox{T = 1250 K} and surface gravity, \hbox{log $g = 1.33$ ms$^{-1}$}. 

In \citet{barnes07hd75289}, the modelled planet, \hbox{HD 75289b}, was not as close to the G0V parent star \hbox{($a = 0.0482$\ AU)} as \hbox{HD 189733b} is to it's K2V host \hbox{($a = 0.0312$\ AU)}. The opacities that remain after rainout of condensed species has occurred (such as TiO and VO) appear stronger for \hbox{HD 189733b} when compared with \hbox{HD 75289b} \citep{barnes07hd75289} and are due to two main species, namely H$_2$O and CO, in the wavelength region of our data. Fig. \ref{fig:planet_spectrum} shows the wavelength range of our data where the tick marks represent strongest H$_2$O, CO and HF opacities. The detection method used to search for a planetary signature (\S \ref{section:decon} and \S \ref{section:matchedfilter}) is model dependent and may be affected by a number of factors. Missing or incorrect line opacities in the model are potential problems; however, the line opacities used in the model are known to be fairly reliable for similar temperatures and pressures found in brown dwarfs \citep{freedman07}. In \citet{barnes07hd75289}, we investigated the effects of a mismatch in line opacity strengths and line positions for a similar instrumental setup to that used for the observations in this paper. We found that the simulations still enabled us to detect a planetary signal with 99.9 per cent confidence (a 1.58 times drop in planet/star ratio) for the case where the model line list line strengths used in deconvolution were modified by 50 per cent or for the case where the 15 percent of the line opacity wavelengths were randomised. The H$_2$O line lists used in our model are taken from \citet{barber06water} and represent the most complete and accurate water line lists to date.}

\subsection{Deconvolution}
\protect\label{section:decon}

{ 
A line list (comprising opacity wavelength positions and line depths) based on the model atmosphere described in \S \ref{section:atmos} is used as a deconvolution template for planetary absorption features which are hidden within each \hbox{HD 189733b} spectrum. The least squares deconvolution procedure yields a single mean absorption line for each \hbox{HD 189733b} spectrum which when scaled with the line list gives the best fit to the observed spectrum. In this way the effective S/N ratio of the mean deconvolved absorption line is much higher than that of individual lines, potentially enabling a weak signal to be extracted from the data. The method of least squares deconvolution was first described by \citet{donati97zdi}. Our implementation of the algorithm \citep{barnes98aper} propagates errors from the input spectra and has been used in reflected light searches in the optical by \citet{cameron99tauboo,cameron02upsand} and \citet{leigh03a,leigh03b}. A detailed description of the use of this algorithm with near infrared spectra is described in \citet{barnes07hd75289}. 

Of the 225 usable spectra, the mean S/N ratio, measured from residual spectra after removal of the template, was \hbox{$136 \pm 85$}. The minimum and maximum S/N ratios were 27 and 382 respectively with only 10 per cent of the spectra possessing S/N $> 250$. We assess the consequences of this highly variable S/N in section \S \ref{section:results}. Since the spectra are dominated by very strong telluric lines in the 2.0445\,-\,2.0610 \micron\ region of order 37 and the 2.3650\,-\,2.3807 \micron\ region of order 32 (Fig. \ref{fig:planet_spectrum}), we did not use these regions in the deconvolution procedure. The gain in S/N of 20 for the mean deconvolved spectrum yielded S/N ratios of \hbox{$2751 \pm 1772$}. The full range of S/N ratios in the deconvolved spectral profiles is however 540\,-\,7640, again, reflecting the highly variable cloud conditions throughout the night of observations.}

\begin{figure*}
 \begin{center}
   \begin{tabular}{cc}

      \vspace{10mm} \\
      \includegraphics[width=70mm,bbllx=113,bblly=114,bburx=397,bbury=397,angle=0]{Ims/barnes_hd189733_fig2.ps} \hspace {5mm} &
      \hspace {5mm}
      \includegraphics[width=70mm,bbllx=113,bblly=114,bburx=397,bbury=397,angle=0]{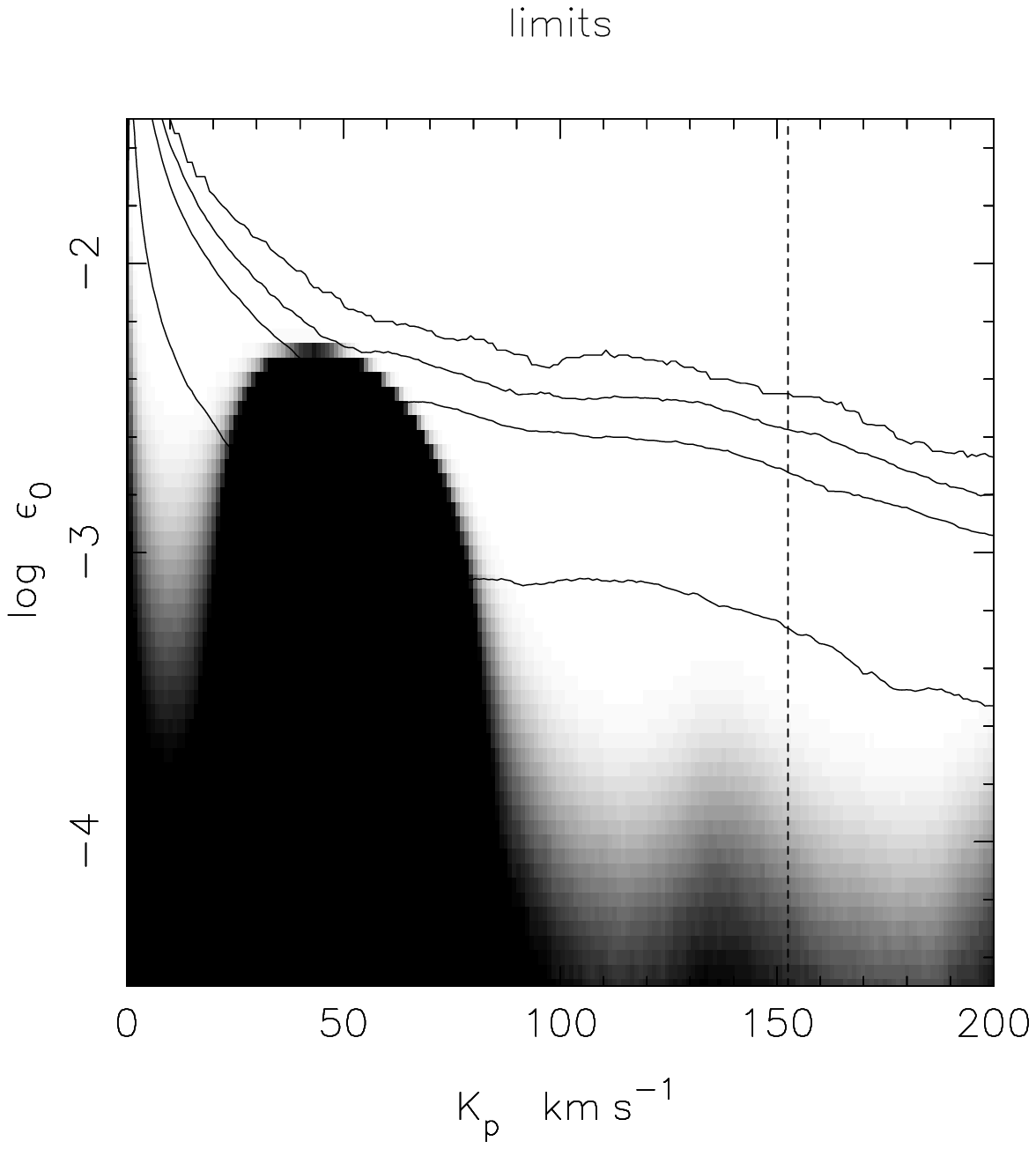} \\

      \vspace{20mm} \\
      \includegraphics[width=70mm,bbllx=113,bblly=114,bburx=397,bbury=397,angle=0]{Ims/barnes_hd189733_fig4.ps} \hspace {5mm} &
      \hspace {5mm}
      \includegraphics[width=70mm,bbllx=113,bblly=114,bburx=397,bbury=397,angle=0]{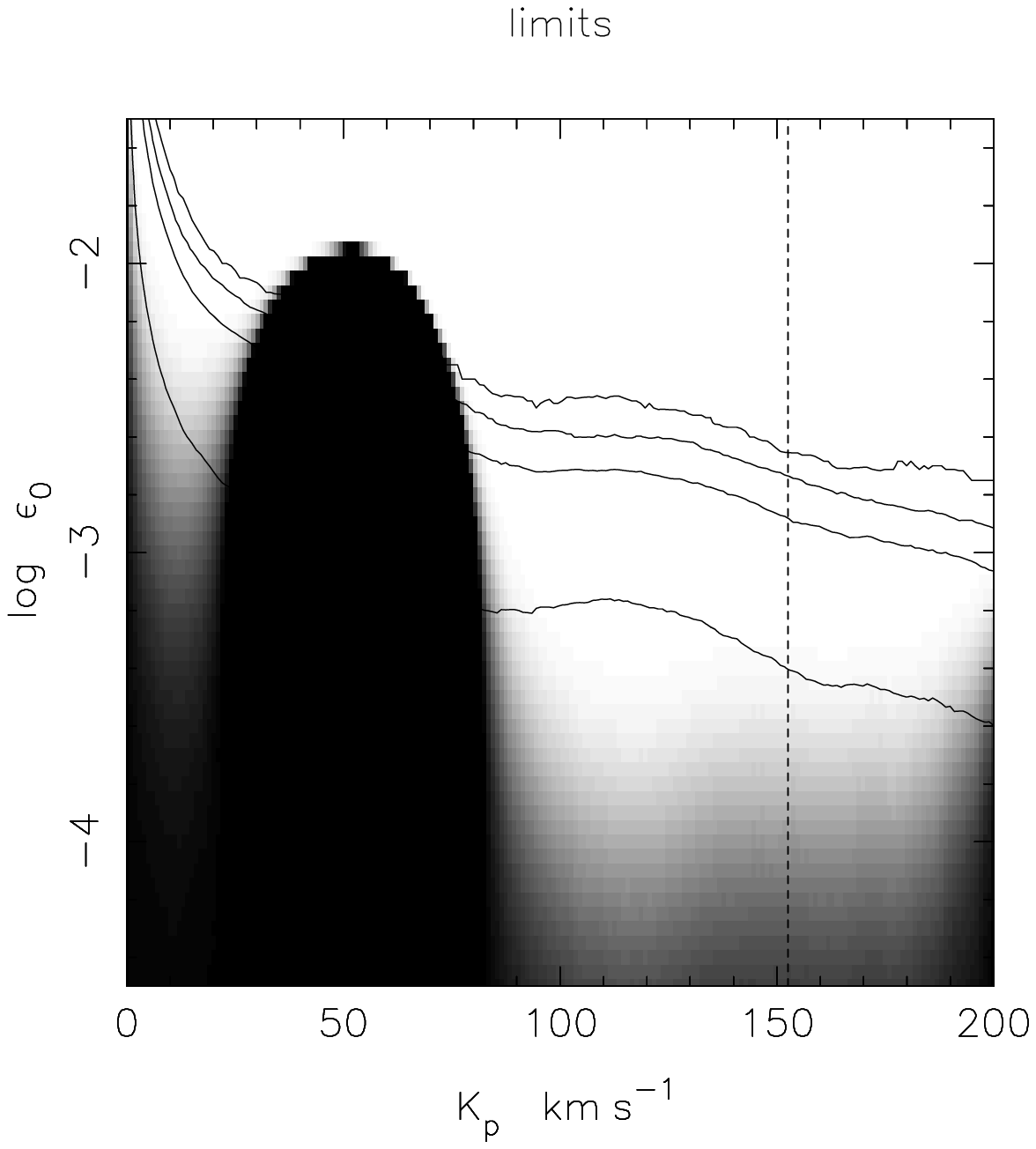} \\

      \vspace{10mm} \\
   \end{tabular}
 \end{center}

\caption{Left: Phased timeseries of the deconvolved residual spectra of \hbox{HD 189733}. The dashed sinusoidal curve represents the {\em expected} motion of a planetary signal based on the empirically determined velocity amplitude, \hbox{$K_p = 152.8 \pm 2.0$}~\kms. Since the S/N ratio is variable, for plotting purposes, all deconvolved profiles have been normalised to the same noise level to optimise the visible information content of the plot. Right: Relative probability \chisq\ map of planet-star flux ratio $log_{10}\,\epsilon_{0}$ vs $K_p$. The first 2 to 5 principal components have been subtracted from the timeseries spectra. The greyscale in the probability maps represents the probability relative to the best fitting model in the range $100$ \kms\ $< K_p \leqslant 200$ \kms\ in the range white for 0 to black \hbox{for 1}. A candidate feature with \hbox{$K_p = 52$ \kms}\ resulting from systematics (see main text) has resulted in the large dark region in the left of the plot. Plotted are the 68.3, 95.4, 99 and 99.9 per cent (bottom to top) confidence levels for a planetary signal detection. The dashed vertical line represents the  velocity amplitude, \hbox{$K_p = 152.8$} \kms. No planetary signal is detected at this velocity. Top: Analysis using all 225 spectra. Bottom: Analysis after removal of blocks of the lowest S/N ratio spectra. Using the remaining 131 deconvolved spectra results in the same confidence levels appearing at lower contrast ratios.}
\protect\label{fig:results}
\end{figure*}

\subsection{Matched filter}
\protect\label{section:matchedfilter}

We model the time-dependent changes in Doppler shift and brightness using a matched filter analysis described in detail in \hbox{Appendix D} of \citet{cameron02upsand}. Modifications of this implementation are described in \citet{barnes07hd75289}, but we reiterate the basics of the procedure here for clarity.

We model a travelling Gaussian which traces out the motion of \hbox{HD 189733b} in an assumed circular orbit about its host star. The radial velocity motion of the planet as a function of phase thus traces out a sinusoid. We assume the planet to be tidally locked and use a model describing the variation in brightness of an irradiated planet that possesses a hot inner face with little redistribution of heat. We assume, as found in \citet{barnes07hd75289}, that this variation closely resembles a Venus-like phase function \citep{hilton92}. Discussion of the true nature of the phase function is discussed in more detail in \S \ref{section:discussion} in light of recent observations, although we note here that a Venus-like phase function agrees with these findings to within 5 per cent over the range of phases of our observations. We asses the relative probabilities of the \chisq\ fits to the data by varying the peak contrast ratio, $\epsilon_{0}(\lambda)$, and the velocity amplitude of the orbital motion, $K_{p}$. The peak contrast ratio occurs at secondary eclipse (i.e. phase \mbox{$\phi = 0.5$}) in the case of a Venus-like phase function. In the phase range \hbox{($0.44 < \phi < 0.59$)} of our data, we are sampling the region of the sinusoid where, to first order, the radial velocity motion approximates a straight line. The gradient of this line, which is fitted by a sinusoid in our model, thus enables an estimate of the velocity amplitude, $K_{p}$, to be made. The improvement in \chisq\ is estimated for combinations of these parameters and is normalised relative to the best-fitting model. To calibrate any candidate detection, we construct a simulated planet signal of known $\epsilon_{0}(\lambda)$, with \hbox{$\lambda = 2.185$\ \micron}\ that is added to the extracted spectra prior to analysis. By ensuring the fake planet is recovered at the correct contrast ratio (hereafter, simply $\epsilon_{0}$ or log$_{10}\,\epsilon_{0}$) by our procedures, we can be confident of a detection in the presence of a genuine planet signal. 

The significance of the result is assessed using bootstrap statistical procedures based on the random re-ordering of the data in a way that scrambles phases while preserving the effects of any correlated systematic errors \citep{cameron02upsand}. The order of the observations is randomised in a set of 3000 trials which will scramble any true planetary signal while enabling the data to retain the ability to produce spurious detections through the chance alignment of systematic errors. The least squares estimate of log$_{10}\,\epsilon_{0}$ and associated \chisq\ as a function of $K_p$ enable us to plot 68.4, 95.4, 99.0 and 99.9 per cent bootstrap limits on the strength of the planetary signal.

\section{Results}
\protect\label{section:results}

In Fig. \ref{fig:results} (left), we plot the deconvolved spectral timeseries. The dashed line is plotted as a guide for the reader and represents the velocity path of a planet with the same parameters as determined empirically for \hbox{HD 189733b}. We can obtain an estimate of the velocity amplitude of the planet by taking a = \mbox{$0.0312 \pm 0.0004$ AU}, $P =$ \mbox{$2.2185733 \pm 0.0000019$\ d} and \mbox{$i = 85.76 \pm 0.29$\degs}, from which we find \mbox{$v_p = 152.58 \pm 1.96$}~\kms. We discuss additional factors that could augment this uncertainty in \S \ref{section:discussion}. Fig. \ref{fig:results} (right) presents the result of our matched Gaussian filter analysis. Again the dashed line represents the velocity at which \hbox{HD 189733b} should appear. The darkest regions in the map represent the greatest improvement in \chisq\ when fitting the model described in \S \ref{section:decon}. In the optical studies carried out by \citet{cameron99tauboo,cameron02upsand,leigh03a,leigh03b}, the reflected light signal was simply an attenuated copy of the stellar spectrum. The central portion of the deconvolved timeseries was excluded during the matched filter modelling to ensure that no spurious signal due to remaining residuals of an inadequately subtracted stellar profile were included. Here however, the planetary spectrum is not expected to be a copy of the stellar spectrum and any line opacities should be randomly distributed with respect to stellar opacities in the spectrum. However, since \hbox{HD 189733b} is a transiting planet, when applying our matched filter analysis, we exclude the portion of spectrum when the planet is not visible. \citet{winn07hd189733b} estimate a transit time of \hbox{$1.827 \pm 0.029$ hrs} indicating that the planet is not visible for phases $0.483 < \phi < 0.517$.

The feature at \hbox{$K_p \sim 52$ \kms} is the result of remaining systematics in the deconvolved timeseries. To enhance the greyscale in the region of the \chisq\ map where we know \hbox{HD 189733b} should appear, as outlined above, we excluded $K_p < 100$ \kms\ when computing the greyscale, resulting in a dark block in the map around the \hbox{$K_p \sim 52$ \kms} feature. We do not detect the planet with 1-$\sigma$ and 2-$\sigma$ confidence at levels of log$_{10}$\,$\epsilon_{0} = {\rm log}_{10}(F_p/F_*) = -3.26$\ and\ $-2.72$ respectively. The equivalent star/planet contrasts are $F_*/F_p = 1820$\ and\ $525$ respectively.  

Fig. \ref{fig:results} (top left) shows clear systematic features, which are not subtracted, even after removing the first 2\,-\,5 principal components as outlined in \S \ref{section:removespec}. This is especially noticeable for spectra taken in the phase range \hbox{$0.52 < \phi < 0.54$}. We attribute these residuals to variation in telluric strengths throughout the night and as a result of the highly variable cloud cover. Since the observations were made at a spectral resolution of $R \sim 15,000$, it is not possible to effectively scale the template spectrum in all regions, especially where those telluric features that may vary in strength independently are blended. 


To investigate the effect of the highly variable S/N, we rejected frames with the lowest S/N ratios. A threshold level that removed the main blocks of lower S/N spectra \hbox{( S/N $< 97$)} was chosen, leaving 131 spectra. The deconvolved profiles (Fig. \ref{fig:results}, bottom left) possess \hbox{S/N = $3800 \pm 1480$}. The matched filter analysis performed on these spectra alone enables us to place 1-$\sigma$ and 2-$\sigma$ confidence at levels of log$_{10}$\,$\epsilon_{0} = -3.40$\ and\ $-2.88$ on a non-detection of \hbox{HD 189733b} (Fig. \ref{fig:results}, bottom right) . The equivalent star/planet contrasts are $F_*/F_p = 2512$\ and\ $759$ respectively. We are nevertheless able to improve the sensitivity of our result, although at the 1-$\sigma$ level, the results are indistinguishable. Despite using fewer spectra, we believe that an increased sensitivity is due to the reduced range of S/N values of our spectra. The systematic effects introduced by unresolved telluric features, which cannot easily be corrected for \citep{bailey07}, are thus a critical factor in determining the sensitivity of our technique.

\section{Discussion}
\protect\label{section:discussion}

The near infrared planet/star contrast ratio, $F_p/F_*$, offers a much more favourable regime over optical wavelengths in which to carry out a search for a planetary signal. The technique, in addition to offering a chance to determine the contrast ratio, relies upon our ability to detect absorption in the planetary atmosphere. The contrast limit placed by this study is intriguing because we do not detect the planet at a level of $F_p/F_*$ that is 1.7 times less than the model prediction; and despite the observation that our model is in agreement with mid-infrared Spitzer observations \citep{deming06hd189733b,grillmair07spectrum,knutson07hd189733b}. Fig. \ref{fig:fluxrat_plot} is a plot of the theoretical $F_p/F_*$ from our model indicating both the mid-infrared Spitzer measurements and the upper limit determined above. { For comparison with the \citet{barnes07hd75289} simulations, a non-detection at the reported level may arise if 55 per cent of line strengths are mis-matched or 16 per cent of the lines possess incorrect wavelengths (see \S \ref{section:atmos}).

\begin{figure}
 \begin{center}
   \begin{tabular}{c}
      \hspace{-3mm}
      \includegraphics[width=86mm,angle=0]{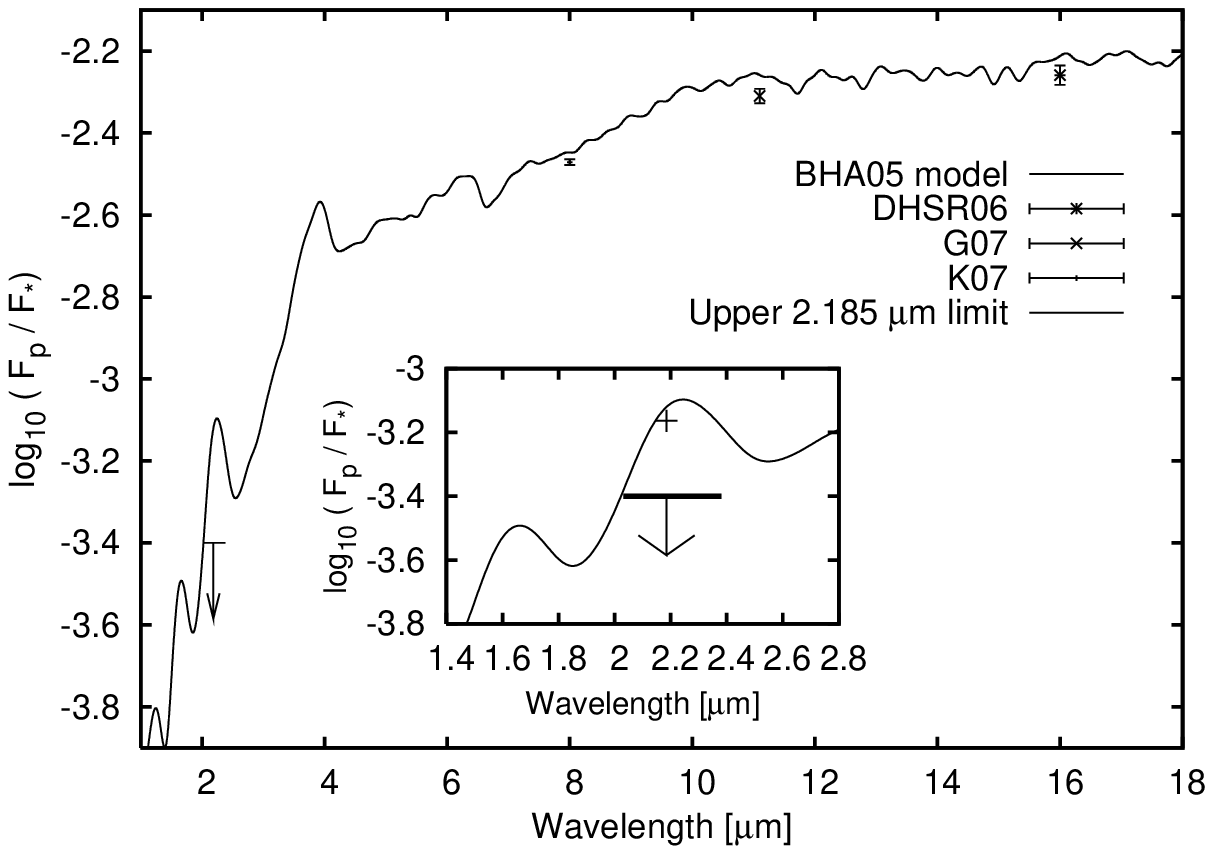} \\

   \end{tabular}
 \end{center}
\caption{Model planet/star flux ratio for the \hbox{HD 189733 system}. A Spitzer eclipse depth measurement using the Infrared Array Camera (IRAC) is plotted for 8 \micron\ (\citet{knutson07hd189733b}, K07). Similar measurements using the Spizter Infrared Spectrograph (IRS) are plotted for 11.1 \micron\ (\citet{grillmair07spectrum}, G07) and 16 \micron\ (\citet{deming06hd189733b}, DHSP06). A horizontal bar with vertical down-pointing arrow indicates our 2.185 \micron\ \hbox{$1-\sigma$} upper limit. The width of the horizontal bar represents the wavelength range of our data. The single point plotted in the inset at log$_{10}$\,$\epsilon_{0} = -3.16$ represents the model mean flux over this range. The upper limit is at a level of log$_{10}$\,$\epsilon_{0} = 0.24$ lower than the model (i.e. $\epsilon_{0}$(model)/$\epsilon_{0}$(observed) = 1.7).}
\protect\label{fig:fluxrat_plot}
\end{figure}

Apart from line position and strength mismatches between model and observed spectra,} another possibility that would result in a non-optimal search for the planetary signature is use of an incorrect phase function. \citet{barnes07hd75289} found a Venus-like phase function for the \hbox{HD 75289} models over the phases of observation considered. The overall variation in flux found by \citet{knutson07hd189733b} indicates that a Venus-like phase function is less appropriate for \hbox{HD 189733b}, indicating that heat is more effectively transported in its atmosphere. The lower contrast between night and day yields a temperature range of 973\,-\,1212 K. Over the small range of our observations however, we find that the variation in flux from the maximum as observed by \citet{knutson07hd189733b} is consistent with a Venus-like phase function to within 5 per cent.

\citet{harrington06} measured the phase dependent flux of $\upsilon$ And b, finding variation that was well fitted by the models of BHA05, implying poor transfer of thermal energy. On the other hand, the more recent observations by \citet{knutson07hd189733b} found that a maximum in flux for the \hbox{HD 189733} system occurred at a phase angle of 16\degs\ before mid-eclipse of the planet. They derived a surface map of \hbox{HD 189733b} and found the hottest part of the atmosphere is offset 30\degs\ East of the substellar point. We adjusted our Venus-like phase function by 16\degs\ to ensure a maximum occurred at the same phase (i.e. $\phi = 0.456$) rather than at phase $\phi = 0.5$. This did not result in the appearance of a candidate planetary feature in our \chisq\ map however. The resulting {\em increase} of the 1-$\sigma$ confidence level from log $\epsilon_0$ = -3.4 to log $\epsilon_0$ = -3.25 is equivalent to a reduction in sensitivity by a factor of 1.4. This is probably the result of shifting the time at which the planetary flux maximum occurs to the phase where observations began. As can be seen in Fig. \ref{fig:results} (bottom left), only one observation is made before maximum when the shifted phase function is used.

Nevertheless, the possibility that the observed phase function peak may be a function of wavelength is supported by simulations. \citet{cooper05} and \citet{cooper06} carried out three-dimensional atmospheric circulation simulations for \hbox{HD 209458b} and found that the circulation pattern is a function of atmospheric pressure or depth. In the lower pressure regime of the outer atmosphere, the radiative timescale is much shorter than advective timescales such that absorbed stellar heat is quickly re-radiated. Lower in the atmosphere, at higher pressures, advective timescales become important and carry absorbed energy in an easterly direction. Based on these models, \citet{fortney06} has shown that the location of the phase function peak is indeed expected to be a function of wavelength in the case of \hbox{HD 209458b}. The time at which the phase function peak occurs before secondary eclipse is inversely proportional to wavelength, in agreement with the findings of \citet{harrington06} and \citet{knutson07hd189733b}. This implies that shorter wavelengths probe deeper layers of the planetary atmosphere. Wind speeds of order $6.6-4.3$ \kms\ at 10 mbar to 1 bar in an easterly direction are high, and while not sufficient to significantly shift the apparent velocity amplitude of the planet (e.g. see Fig. \ref{fig:results}), they introduce an additional factor into the error budget of the estimated 1.96 \kms\ uncertainty (\S \ref{section:results}). An easterly wind flow would act to reduce the apparent velocity amplitude, but since most of the planetary signal is expected to come from the hotspot which would be travelling transversely to the observer at the observed phases, any induced perturbation is likely to be a second order effect.

Although \citet{fortney06} do not consider wavelengths short of 3.6 \micron\ it seems reasonable to expect that our $\sim2.2$ \micron\ observations also probe deeper layers of the atmosphere. If the temperature-pressure gradient is lower or even flat in this regime (i.e. 100mbar - 1 bar), as indicated for \hbox{HD 209458b} in BHA05, it may be that absorption lines are suppressed and the dayside spectrum approaches that of a black body. Also, the reasonable agreement between the hotter no-redistribution model and the Spitzer observations (that emerge from layers above the $K$-band photosphere), may be an indication of a deeper temperature structure cooled by energy redistribution to the night side, similar to what has been predicted for \hbox{HD 209458b} by \citet{cooper05}. Constraints on the depth-dependent redistribution in the atmosphere of \hbox{HD 189733b} will be explored in more detail by Barman et al. (in prep.). 


\section{Conclusion}
\protect\label{section:conclusion}

The methods presented here provide a powerful technique with which to detect a planetary absorption signature and the 1-$\sigma$ log$_{10}$\,$\epsilon_{0}$ = -3.40 limit is extremely interesting. We have outlined a number of possibilities for the lack of detection of a planetary signal. Our data suggest that ideal observing conditions would enable us to obtain spectra with S/N ratios as high as $\sim 380$. With the same NIRSPEC instrumental setup at Keck II, a total of 230 frames with S/N = 300 taken over the same orbital phase range would enable us to push our $1-\sigma$ and $2-\sigma$ limits down to levels of \hbox{log$_{10}$\,$\epsilon_{0}$ = -3.90 \& 3.40} respectively. Also, the high variability of the stellar flux relative to the telluric lines (due to highly variable cloud cover) seen in our present data would be reduced in more ideal conditions. This would lead to fewer systematics after template spectrum removal enabling us to further lower our confidence level estimates for $\epsilon_0$. Additionally, observations at higher spectral resolution than currently afforded by the present data would be desirable, as demonstrated in \cite{barnes07hd75289}, and would lead to better telluric subtraction.

An independent, near infrared, photometric detection would enable us to place stronger limits on the reliability of the line list parameters or the nature of the atmosphere itself. For instance, when combined with spectral observations, as described above, it would be possible to rule out or place limits on the presence of absorption features in the corresponding wavelength region.

\section{Acknowledgments}
JRB was supported by a PPARC funded research grant during the course of this work. TB acknowledges support from NASA's Origins of Solar System program and the NASA Advanced Supercomputing facility, and LP from NSF grant 04-44017. The data presented herein were obtained at the W.M. Keck Observatory, which is operated as a scientific partnership among the California Institute of Technology, the University of California and the National Aeronautics and Space Administration. The Observatory was made possible by the generous financial support of the W.M. Keck Foundation. The authors wish to recognise and acknowledge the very significant cultural role and reverence that the summit of Mauna Kea has always had within the indigenous Hawaiian community.  We are most fortunate to have the opportunity to conduct observations from this mountain.


\end{document}